\documentstyle[times,graphicx,mathrsfs]{mn}

\newif\ifAMStwofonts
\AMStwofontstrue

\ifoldfss
  \ifCUPmtlplainloaded \else
    \NewTextAlphabet{textbfit} {cmbxti10} {}
    \NewTextAlphabet{textbfss} {cmssbx10} {}
    \NewMathAlphabet{mathbfit} {cmbxti10} {} 
    \NewMathAlphabet{mathbfss} {cmssbx10} {} 
  \fi
  \ifAMStwofonts
    \ifCUPmtlplainloaded \else
      \NewSymbolFont{upmath} {eurm10}
      \NewSymbolFont{AMSa} {msam10}
      \NewMathSymbol{\upi}     {0}{upmath}{19}
      \NewMathSymbol{\umu}     {0}{upmath}{16}
      \NewMathSymbol{\upartial}{0}{upmath}{40}
      \NewMathSymbol{\leqslant}{3}{AMSa}{36}
      \NewMathSymbol{\geqslant}{3}{AMSa}{3E}

    \fi
  \fi
\fi 

\ifnfssone
  \newmathalphabet{\mathit}
  \addtoversion{normal}{\mathit}{cmr}{m}{it}
  \addtoversion{bold}{\mathit}{cmr}{bx}{it}
  \newmathalphabet{\mathbfit} 
  \addtoversion{normal}{\mathbfit}{cmr}{bx}{it}
  \addtoversion{bold}{\mathbfit}{cmr}{bx}{it}
  \newmathalphabet{\mathbfss} 
  \addtoversion{normal}{\mathbfss}{cmss}{bx}{n}
  \addtoversion{bold}{\mathbfss}{cmss}{bx}{n}
  \ifAMStwofonts
    \ifCUPmtlplainloaded \else
      \UseAMStwoboldmath
      \makeatletter
      \new@mathgroup\upmath@group
      \define@mathgroup\mv@normal\upmath@group{eur}{m}{n}
      \define@mathgroup\mv@bold\upmath@group{eur}{b}{n}
      \edef\UPM{\hexnumber\upmath@group}
      \new@mathgroup\amsa@group
      \define@mathgroup\mv@normal\amsa@group{msa}{m}{n}
      \define@mathgroup\mv@bold\amsa@group{msa}{m}{n}
      \edef\AMSa{\hexnumber\amsa@group}
      \makeatother
      \mathchardef\upi="0\UPM19
      \mathchardef\umu="0\UPM16
      \mathchardef\upartial="0\UPM40
      \mathchardef\leqslant="3\AMSa36
      \mathchardef\geqslant="3\AMSa3E
    \fi
  \fi
\fi 

\ifnfsstwo
  \DeclareMathAlphabet{\mathbfit}{OT1}{cmr}{bx}{it}
  \SetMathAlphabet\mathbfit{bold}{OT1}{cmr}{bx}{it}
  \DeclareMathAlphabet{\mathbfss}{OT1}{cmss}{bx}{n}
  \SetMathAlphabet\mathbfss{bold}{OT1}{cmss}{bx}{n}
  \ifAMStwofonts
    \ifCUPmtlplainloaded \else
      \DeclareSymbolFont{UPM}{U}{eur}{m}{n}
      \SetSymbolFont{UPM}{bold}{U}{eur}{b}{n}
      \DeclareSymbolFont{AMSa}{U}{msa}{m}{n}
      \DeclareMathSymbol{\upi}{0}{UPM}{"19}
      \DeclareMathSymbol{\umu}{0}{UPM}{"16}
      \DeclareMathSymbol{\upartial}{0}{UPM}{"40}
      \DeclareMathSymbol{\leqslant}{3}{AMSa}{"36}
      \DeclareMathSymbol{\geqslant}{3}{AMSa}{"3E}
    \fi
  \fi
\fi 

\ifCUPmtlplainloaded \else
  \ifAMStwofonts \else 
    \def\upi{\pi}
    \def\umu{\mu}
    \def\upartial{\partial}
  \fi
\fi

   \title[On the origin of helium-rich populations]{On the origin of the 
     helium-rich population in ${\bmath{\omega}}$\,Centauri}
   \author[D. Romano et al.]{D.~Romano$^{1, 2}$\thanks{E-mail: 
       donatella.romano@oabo.inaf.it}, M.~Tosi$^{2}$, M.~Cignoni$^{1, 2}$, 
     F.~Matteucci$^{3}$, E.~Pancino$^{2}$ and M.~Bellazzini$^{2}$\\
           $^{1}$Dipartimento di Astronomia, Universit\`a di Bologna,
                 Via Ranzani 1, I-40127 Bologna, Italy\\
           $^{2}$INAF\,--\,Osservatorio Astronomico di Bologna,
                 Via Ranzani 1, I-40127 Bologna, Italy\\
           $^{3}$Dipartimento di Astronomia, Universit\`a di Trieste,
                 Via Tiepolo 11, I-34143 Trieste, Italy}
   
\begin{document}

     \date{Accepted 2009 October 2. Received 2009 October 2; in original form 
       2009 July 31}

     \pagerange{\pageref{firstpage}--\pageref{lastpage}} \pubyear{2009}

     \maketitle

     \label{firstpage}


   \begin{abstract}
     To study the possible origin of the huge helium enrichment attributed to 
     the stars on the blue main sequence of $\omega$\,Centauri, we make use of 
     a chemical evolution model that has proven able to reproduce other major 
     observed properties of the cluster, namely, its stellar metallicity 
     distribution function, age-metallicity relation and trends of several 
     abundance ratios with metallicity. In this framework, the key condition to 
     satisfy all the available observational constraints is that a 
     galactic-scale outflow develops in a much more massive parent system, as a 
     consequence of multiple supernova explosions in a shallow potential well. 
     This galactic wind must carry out preferentially the metals produced by 
     explosive nucleosynthesis in supernovae, whereas elements restored to the 
     interstellar medium through low-energy stellar winds by both asymptotic 
     giant branch (AGB) stars and massive stars must be mostly retained. 
     Assuming that helium is ejected through slow winds by both AGB stars and 
     fast rotating massive stars (FRMSs), the interstellar medium of 
     $\omega$\,Centauri's parent galaxy gets naturally enriched in helium in 
     the course of its evolution.
   \end{abstract}



   \section{Introduction}
   \label{sec:int}

   Once referred to as `simple stellar populations', globular clusters (GCs) 
   are not simple at all, in that they display significant and peculiar 
   star-to-star abundance variations (see Gratton, Sneden \& Carretta 2004 for 
   a review). These are most likely the records of a complex star formation 
   history. Indeed, the C-N, Na-O, Al-O and Al-Mg anticorrelations seen among 
   both evolved and unevolved stars of individual clusters cannot be reconciled 
   with a simple deep mixing scenario (Gratton et al. 2001; Cohen, Briley \& 
   Stetson 2002). Rather, they are naturally explained if one or more 
   successive stellar generations form out of the ejecta from first generation 
   stars that have undergone nuclear processing through proton capture 
   reactions at high temperatures (Sneden et al. 2004; Carretta et al. 2006, 
   and references therein). This evolutive picture is supported by the presence 
   of multiple main sequences (MSs) in two massive Galactic GCs, $\omega$\,Cen 
   and NGC\,2808 (Bedin et al. 2004; Piotto et al. 2007). The bluer MSs are -- 
   surprisingly -- more metal-rich than the red ones, or have the same iron 
   content. This, at present, can be understood only in terms of an extreme 
   helium enhancement in the blue population (Norris 2004; D'Antona et al. 
   2005; Piotto et al. 2005). Further evidence in favour of the existence of 
   very helium-rich subpopulations in massive clusters comes from the peculiar 
   morphology of the horizontal branches in some of them (Busso et al. 2007; 
   Caloi \& D'Antona 2007; Yoon et al. 2008; but see also Catelan et al. 2009). 
   The excess in the far-UV flux detected for most of the massive clusters 
   observed in M\,87 by Kaviraj et al. (2007) can also be interpreted as a 
   signature of extreme helium enrichment and would demonstrate that this 
   phenomenon is not limited to our Galaxy.

   It is common wisdom that key to understanding the origin of extreme-helium 
   populations is the identification of extreme helium polluters. Though 
   massive (approximately 5--10 M$_\odot$) asymptotic giant branch (AGB) stars 
   have been suggested as likely `culprits' for the chemical `anomalies' 
   observed in GC stars since the work by Cottrell \& Da Costa (1981), only 
   recently detailed stellar modelling has provided yields compatible with the 
   observations over a significant range of initial stellar masses (Pumo, 
   D'Antona \& Ventura 2008; Ventura \& D'Antona 2008a,b). In Ventura \& 
   D'Antona's models, convection is modelled efficiently and a very fast AGB 
   evolution is obtained, which results in a relatively low number of thermal 
   pulses (TPs) and third dredge-up (TDU) episodes. This leads to a lower 
   production of carbon and nitrogen and no appreciable increase of the overall 
   CNO abundance in the envelope, an important point in view of the fact that 
   in $\omega$\,Cen and other globulars the [C+N+O/Fe]\footnote{Unless 
     otherwise stated, chemical abundances throughout this paper are by number, 
     except for $Y$ and $Z$, that indicate the mass fraction of helium and 
     total metals, respectively. As usual, log\,$\varepsilon$(X)~$\equiv$ 
     12\,+\,log\,(X/H), while square brackets indicate logarithmic ratios 
     relative to solar, [A/B]~$\equiv$ log\,(A/B)\,$-$\,log\,(A/B)$_{\odot}$.} 
   ratio is constant within a factor of about 2 (Norris \& Da Costa 1995; Smith 
   et al. 1996; Ivans et al. 1999; Carretta et al. 2005; Cohen \& Mel\'endez 
   2005; but see Yong et al. 2009).

   Up to now, chemical evolution models aimed at explaining the large helium 
   abundance implied by the blue MS of $\omega$\,Cen with the ejecta of AGB 
   stars have predicted an enormous increase in the total C+N+O content of the 
   cluster, contrary to observations (e.g. Karakas et al. 2006). This failure 
   has spurred the quest for alternative solutions. Decressin et al. (2007a,b) 
   have proposed a scenario where the H-processed material lost by first 
   generation fast rotating massive stars (FRMSs) through slow mechanical 
   equatorial winds is retained in the cluster potential well, where it enters 
   the formation of second generation stars. Matter expelled through fast polar 
   winds and supernova (SN) explosions leaves the cluster instead. A major 
   drawback with this scenario is that it cannot estimate from first principles 
   the efficiency of meridional circulation to mix helium (and other chemicals) 
   into the stellar envelope, nor the rate of mass loss through the outflowing 
   disc (Renzini 2008). Furthermore, the helium-rich stars would form in very 
   `hostile' surroundings, their birth being shortly followed by -- or even 
   concomitant to -- multiple SN explosions.

   To reproduce the correct ratio of first generation-to-second generation 
   stars in both the massive star pollution scenario and the AGB pollution 
   scenario, either a highly anomalous initial mass function (IMF) for first 
   generation stars or a strong evaporation of low-mass stars with `normal' 
   chemical enrichment have to be assumed (Decressin et al. 2007b; D'Ercole et 
   al. 2008). While, in general, the first condition is difficult to justify 
   both theoretically and observationally, in the case of $\omega$\,Cen there 
   are fairly clear indications that the second generation of stars could have 
   formed from the ejecta of surrounding field stellar populations in an 
   initially much more massive system. The kinematical, dynamical and chemical 
   properties of this cluster, in fact, are better understood if it is the 
   surviving nucleus of an ancient dwarf galaxy captured and disrupted by the 
   gravitational field of the Galaxy many Gyr ago (see, for instance, Dinescu, 
   Girard \& van Altena 1999; Gnedin et al. 2002; Bekki \& Norris 2006; Romano 
   et al. 2007; Bellazzini et al. 2008). Recent {\it N}-body simulations of the 
   dynamical evolution of two-population clusters show that a rapid loss of 
   first generation stars should be expected early on in the cluster evolution 
   as a consequence of cluster expansion in response to the dynamical heating 
   from SN explosions (D'Ercole et al. 2008; see also Decressin, Baumgardt \& 
   Kroupa 2008). Early phases of violent relaxation also sensibly reduce the 
   initial cluster's mass (Meylan \& Heggie 1997, and references therein). 
   $\omega$\,Cen, however, must have followed a different evolutionary path, 
   with its putative progenitor galaxy sheding stars in trails while its orbit 
   was degrading by approaching the Milky Way plane (Meza et al. 2005). The 
   debris of this disruption process has been possibly identified through the 
   kinematical feature imprinted in a sample of local, metal-poor halo stars 
   (Dinescu 2002; Mizutani, Chiba \& Sakamoto 2003).

   In this paper, we deal with the chemical evolution of the system that once 
   was/contained $\omega$\,Cen. In particular, in Section~\ref{sec:omega} we 
   propose an explanation for the origin of its helium-rich population, in the 
   context of a model that has already proven able to reproduce the majority of 
   the observational constraints available for its whole, complex stellar 
   population. For sake of completeness, in Section~\ref{sec:omega} we also 
   examine, in our framework, the two alternative scenarios of He being 
   overenriched either only by AGBs or by FRMSs. Our conclusions are presented 
   in Section~\ref{sec:end}, together with a discussion of the results.

   \section{The chemical evolution of $\bmath{\omega}$\,Centauri's 
     progenitor system}
   \label{sec:omega}

   \subsection{The chemical evolution model}
   \label{sec:code}

   In this paper we adopt an updated version of the chemical evolution model 
   developed by Romano et al. (2007) for a dwarf spheroidal galaxy whose dense 
   central regions become $\omega$\,Cen after accretion and stripping by the 
   Milky Way. Here we simply recall the overall evolutionary scenario, while a 
   detailed description of the model basic assumptions and equations is given 
   in the Appendix.

   We start our computation with $\mathscr{M}_{\mathrm{gas}}$($t$~= 0)~= 
   $\mathscr{M}_{\mathrm{bar}}$~= 10$^9$ M$_\odot$ of gas of primordial chemical 
   composition available for accretion, embedded in a 10 times more massive 
   dark matter halo. The system accretes gas and forms stars for 3 Gyr. About 
   200 Myr after the onset of star formation, as a consequence of energy 
   injection by multiple SN explosions, a galactic wind develops and gradually 
   cleans up the cluster of its gas. Star formation may proceed if the SN 
   ejecta are vented out along preferential directions, leaving part of the 
   pristine gas unperturbed (e.g. Recchi et al. 2001), and we assume this is 
   the case. Feeding the system by continuous infall of gas from the outskirts 
   contrasts the cleaning action of the wind. At the end of the computation, we 
   are left with $\mathscr{M}_{\mathrm{stars}}$($t$~= 3 
   Gyr)~$\simeq$10$^8$~M$_\odot$. According to the computations by Bekki \& 
   Freeman (2003), this is exactly what is needed to leave behind a compact 
   remnant of mass $\mathscr{M}_{\omega\,\mathrm{Cen}} \simeq$ 10$^6$~M$_{\odot}$ 
   after long-term tidal interactions with the Milky Way. During the time (from 
   $t$~= 0.6~Gyr to $t$~= 1~Gyr in our model) the metallicity of 
   $\omega$\,Cen's parent galaxy is growing from [Fe/H]~$\simeq -$1.3 to 
   [Fe/H]~$\simeq -$1.1 (that is, the metallicity range where He-rich stars are 
   observed), the system is forming stars at an average rate of 0.1~M$_\odot$ 
   yr$^{-1}$. Therefore, according to the assumed IMF (Salpeter 1955, 
   extrapolated to the 0.1--100 M$_{\odot}$ stellar mass range; see the 
   Appendix), in principle up to 2~$\times$ 10$^7$~M$_{\odot}$ of stars (20 per 
   cent of the overall population) can form in the range 0.1--0.8~M$_\odot$ 
   (i.e. they are still alive today) out of gas enriched in He at a level 
   comparable to that required by the blue MS stars. The percentage of He-rich 
   stars could be even higher, should the processed gas gradually collect into 
   the galaxy's innermost regions, where the newborn stellar generations are 
   likely to be less severely affected by the subsequent interactions with the 
   Milky Way.

   The problem we left unsettled in our previous work is how to reach the high 
   level of He enhancement required by the blue MS stars. We tackle this 
   subject in the present study.

   \subsection{Nucleosynthesis prescriptions}
   \label{sec:nucpr}

   In this work we adopt the following sets of metallicity-dependent yields for 
   single stars:
   \begin{enumerate}
     \item Model A: yields from van den Hoek \& Groenewegen (1997) for low- 
       and intermediate-mass stars (LIMSs) and Woosley \& Weaver (1995) for 
       massive stars;
     \item Model B: yields from Marigo (2001) for LIMSs, Portinari, Chiosi \& 
       Bressan (1998) for quasi-massive stars and Kobayashi et al. (2006) for 
       massive stars, except for He and CNO production, for which the 
       pre-supernova yields from the Geneva group\footnote{Meynet \& Maeder 
         (2002) for $Z_{\mathrm{ini}}$~= 10$^{-5}$ and 0.004, Hirschi, Meynet 
         \& Maeder (2005) for solar initial metallicity, Hirschi (2007) for 
         $Z_{\mathrm{ini}}$~= 10$^{-8}$ and Ekstr\"om et al. (2008) for 
         zero-metallicity stars.} are adopted;
     \item Model C: yields from Karakas \& Lattanzio (2007) for LIMSs and 
       Kobayashi et al. (2006) for massive stars, except for He and CNO 
       production, for which the pre-supernova yields from the Geneva group are 
       adopted;
   \end{enumerate}
   A thorough discussion of the adopted yields can be found in the source 
   papers.

   We also run models adopting the yields by Ventura \& D'Antona (2008a,b; see 
   Sect.~\ref{sec:heAGB}) and Decressin et al. (2007a; see 
   Sect.~\ref{sec:heMASS}). Ventura \& D'Antona's yields are computed for a 
   reduced grid of stellar masses (only stars with initial masses between 3 and 
   6.3 M$_\odot$ are considered). Thus, below 3 M$_\odot$ we couple them with 
   those from other studies (Karakas \& Lattanzio 2007). Decressin et al. 
   (2007a) provide the abundances of Na in the slow winds of FRMSs, but only 
   for stars with initial metallicity [Fe/H]~= $-$1.5. The adoption of 
   metallicity-dependent yields of Na from FRMSs is expected to change both 
   \emph{qualitatively} and \emph{quantitatively} the model predictions 
   presented in this paper, as discussed in Sect.~\ref{sec:heMASS}.

   In the range from 6--8 M$_{\odot}$ (depending on stellar models) to 
   11--12~M$_{\odot}$, detailed stellar yields are not available. Hence, we 
   interpolate among the yields for LIMSs and massive stars listed above. The 
   chemical imprint of stars in this mass range, the so-called super-AGB stars, 
   should mainly affect the model predictions regarding $^4$He (Pumo et al. 
   2008) $^{13}$C, $^{14}$N, $^{25}$Mg, $^{26}$Al and $^{23}$Na evolution (Siess 
   2007). However, in the absence of detailed nucleosynthesis computations, it 
   can hardly be said which effects must be expected.

   For SNeIa we adopt the yields from Iwamoto et al. (1999).

   \subsection{Model results}

   In previous work (Romano et al. 2007; Romano \& Matteucci 2007) we presented 
   a chemical evolution model able to reproduce the main chemical properties of 
   the complex stellar population of $\omega$\,Cen, namely, its stellar 
   metallicity distribution function, age--metallicity relation and the trends 
   of several abundance ratios with metallicity. A key assumption of the model 
   was that the cluster is the compact remnant of a dwarf galaxy that 
   self-enriched over a period of about 3 Gyr, before being captured and partly 
   disrupted by the Milky Way. However, the issue of the extreme helium 
   enhancement of the stars on the blue MS was unsolved. In fact, our 
   homogeneous chemical evolution model, adopting a standard IMF and standard 
   stellar yields, could not predict any significant helium enrichment during 
   the evolution of the cluster (see figure~7 of Romano et al. 2007, and 
   discussion therein).

   Here, we propose an explanation for the origin of the helium-rich population 
   in $\omega$\,Cen in the framework of our chemical evolution model. We 
   examine separately three key scenarios in the following subsections.


   \begin{figure}
     \includegraphics[width=\columnwidth]{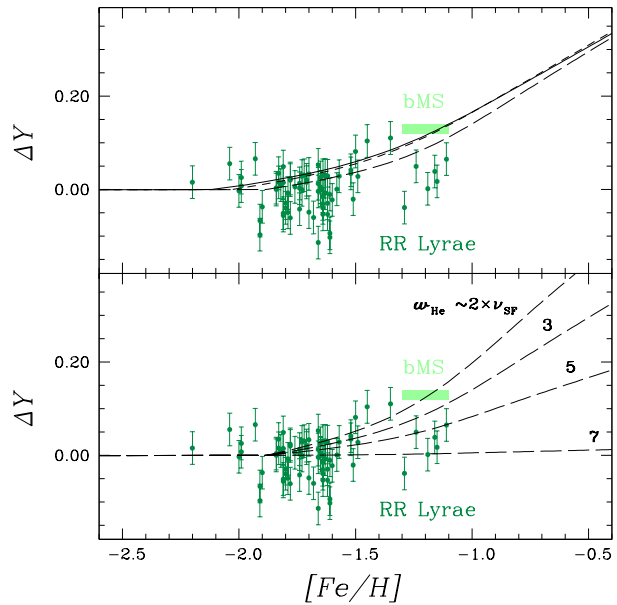}
     \caption{ The He enhancement as a function of metallicity predicted by 
     different models for $\omega$\,Cen is compared to estimates of relative He 
     abundances from RR Lyr\ae \ (small filled circles; Sollima et al. 2006) 
     and blue MS stars (box; Norris 2004; Piotto et al. 2005). Top panel: 
     theoretical predictions from Model A (solid line), Model B (short-dashed 
     line) and Model C (long-dashed line), adopting different nucleosynthesis 
     prescriptions. Bottom panel: theoretical predictions from Model C for 
     different choices of the efficiency of He entrainment in the galactic 
     outflow, as labeled (see text for discussion).}
     \label{fig:dyOmega}
   \end{figure}


   \subsubsection{Differential wind with helium retention}
   \label{sec:heALL}

   In Fig.~\ref{fig:dyOmega}, top panel, we show how the relative He enrichment 
   (assuming $Y_{\mathrm{P}}$~= 0.248 for the primordial He abundance) proceeds 
   in the ISM of our model as a function of metallicity, according to different 
   nucleosynthesis prescriptions (solid line: Model A, short-dashed line: Model 
   B, long-dashed line: Model C; see Sect.~\ref{sec:nucpr}). The efficiency of 
   He ejection through the outflow is set to a low value, $w_{\mathrm{He}} 
   \simeq$3\,$\nu$, about three times the star formation efficiency, in order 
   to reproduce the high He content of the blue MS stars (box in 
   Fig.~\ref{fig:dyOmega}). It can be seen that changing the nucleosynthesis 
   prescriptions does not alter much the model predictions. 

   On the other hand, any modification in the efficiency of He removal from the 
   system dramatically affects the results. In Fig.~\ref{fig:dyOmega}, bottom 
   panel, we show the behaviour of the relative He enrichment for Model C (but 
   the results would be qualitatively the same for Models~A and B), with four 
   different choices for the efficiency of He entrainment in the galactic wind. 
   It is worth noticing that, if different zones of the proto-cluster lose 
   their He with varying strength in the outflow, a spread in the abundance of 
   He results. This could explain the coexistence of populations with `normal' 
   and `enhanced' He abundances at the same metallicity, as seems to be 
   required by observations of RR~Lyr\ae \ stars (Sollima et al. 2006; small 
   filled circles in Fig.~\ref{fig:dyOmega}).

   Thus, lowering the efficiency of He ejection through the outflow is a 
   promising way to get He-enhanced stellar populations. However, it is 
   mandatory to test the proposed scenario against other observed quantities. 
   In particular, consistency between model predictions and observations has 
   to be obtained for other species produced in lockstep with He, which should 
   share the same fate.


   \begin{figure}
     \includegraphics[width=\columnwidth]{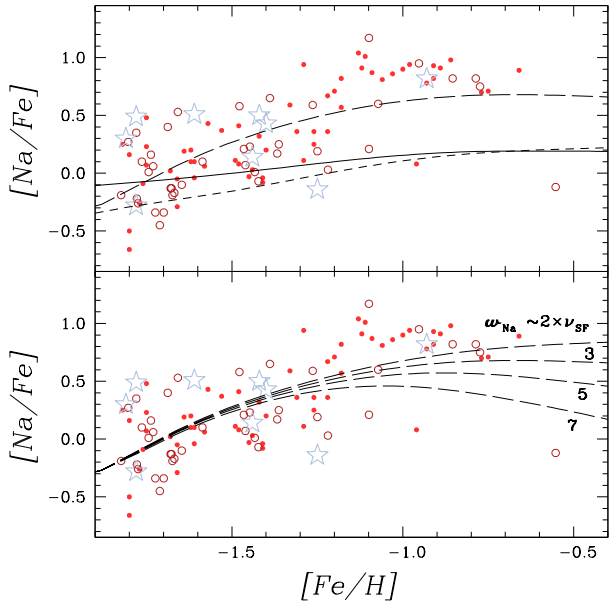}
     \caption{ [Na/Fe] versus [Fe/H] in $\omega$\,Cen. Data are from Norris \& 
       Da Costa (1995; empty circles), Smith et al. (2000; stars) and Johnson 
       et al. (2009; small filled circles). Top panel: the solid, short-dashed 
       and long-dashed lines are for Model A, B and C predictions, 
       respectively. Bottom panel: Model C predictions, for different choices 
       of the efficiency of Na ejection through the outflow (see text).}
     \label{fig:naOmega}
   \end{figure}


   Sodium is produced by stars across the whole mass range, like He. It is 
   synthesized partly during hydrostatic carbon burning, but mostly during 
   hydrogen burning (through the NeNa cycle) in the envelopes of AGB stars and 
   in the cores of massive stars. According to Decressin et al. (2007a), in 
   massive fast rotators rotational mixing efficiently transports elements from 
   the convective cores to the surfaces. If the initial rotational velocities 
   are high, the stars reach the break-up early on the main sequence and eject 
   important quantities of material loaded with H-burning products. Therefore, 
   the fraction of Na produced by massive fast rotators is restored to the ISM 
   by low-energy stellar winds, rather than through SNII explosions, and, thus, 
   follows the same conditions as He for the entrainment in the galactic wind. 
   In Fig.~\ref{fig:naOmega}, we show the predictions on the [Na/Fe] versus 
   [Fe/H] trend in $\omega$\,Cen obtained with different nucleosynthesis 
   prescriptions (top panel) and different efficiencies of Na entrainment in 
   the galactic wind (bottom panel). Models A and B (solid and short-dashed 
   lines, respectively) do not consider Na production from LIMSs. The 
   differences in the model predictions are thus only due to differences in the 
   adopted yields of Na from massive stars (Woosley \& Weaver 1995 for Model~A 
   and Kobayashi et al. 2006 for Model B). Model C (long-dashed lines), 
   instead, includes Na production from LIMSs, through the adoption of Karakas 
   \& Lattanzio's (2007) yields. It has been computed with four values of the 
   efficiency of Na ejection through the outflow, $w_{\mathrm{Na}} \simeq$2, 3, 
   5 and 7 times the efficiency of star formation (Fig.~\ref{fig:naOmega}, 
   bottom panel), the same as for He (Fig.~\ref{fig:dyOmega}, bottom panel). 
   The theoretical predictions are compared to data from Norris \& Da Costa 
   (1995; empty circles), Smith et al. (2000; stars) and Johnson et al. (2009; 
   small filled circles). While no attempt is made to homogenize the data, we 
   note that Johnson et al. (2009) find negligible differences in measured 
   [Fe/H] and [Na/Fe] ratios for 7 stars they have in common with Norris \& Da 
   Costa (1995). From a comparison between model predictions and observations, 
   we conclude that:
   \begin{enumerate}
     \item Model C, accounting for Na production from both LIMSs and massive 
       stars, is able to reproduce the trend of increasing 
       $\langle$[Na/Fe]$\rangle$ with increasing metallicity traced by the 
       majority of the cluster stars, provided that a low efficiency of Na 
       entrainment in the outflow is assumed. This efficiency turns out to be 
       the same required to produce the extreme He-rich population hosted on 
       the blue MS.
     \item A minority of the stars in the intermediate- and high-metallicity 
       domain have Na abundances consistent with production solely from massive 
       stars ([Na/Fe]~$\simeq$0.0; Fig.~\ref{fig:naOmega}, top panel, solid and 
       short-dashed lines). Alternatively, these stars might have formed in 
       regions where Na was more efficiently removed from the ISM 
       (Fig.~\ref{fig:naOmega}, bottom panel, lower long-dashed curve); in that 
       case, the stars should also have `normal' He abundances 
       (Fig.~\ref{fig:dyOmega}, bottom panel, lower long-dashed curve). 
       Indeed, Villanova, Piotto \& Gratton (2009) have recently provided the 
       first direct measurements of the abundance of He for a sample of 5 stars 
       in the Galactic GC NGC\,6752. The stars likely display the original He 
       content of the gas out of which they were born. A mean value of $\langle 
       Y \rangle$~= 0.245 $\pm$ 0.012 is found, consistent with the primordial 
       one. At a metallicity of [Fe/H]~= $-$1.56~$\pm$ 0.03, all these 
       He-normal stars are also Na-poor.
   \end{enumerate}


   \begin{figure}
     \includegraphics[width=\columnwidth]{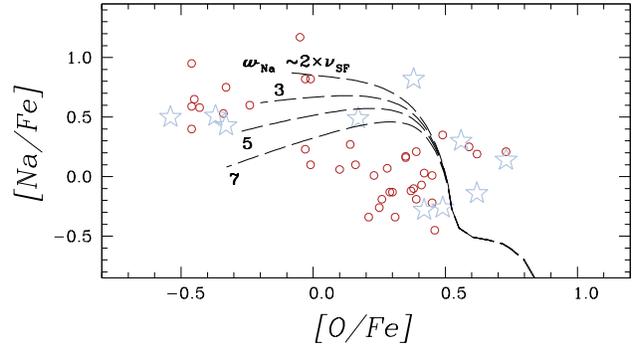}
     \caption{ The Na-O anticorrelation in $\omega$\,Cen. Data are from Norris 
     \& Da Costa (1995; empty circles) and Smith et al. (2000; stars). The 
     predictions of Model C are also shown, for four different values of the 
     efficiency of Na entrainment in the galactic-scale outflow, as labeled. At 
     variance with Figs.~\ref{fig:dyOmega} and \ref{fig:naOmega}, the time on 
     the {\it x}-axis now flows from right to left.}
     \label{fig:anticorr}
   \end{figure}



   \begin{figure*}
     \includegraphics[width=\textwidth]{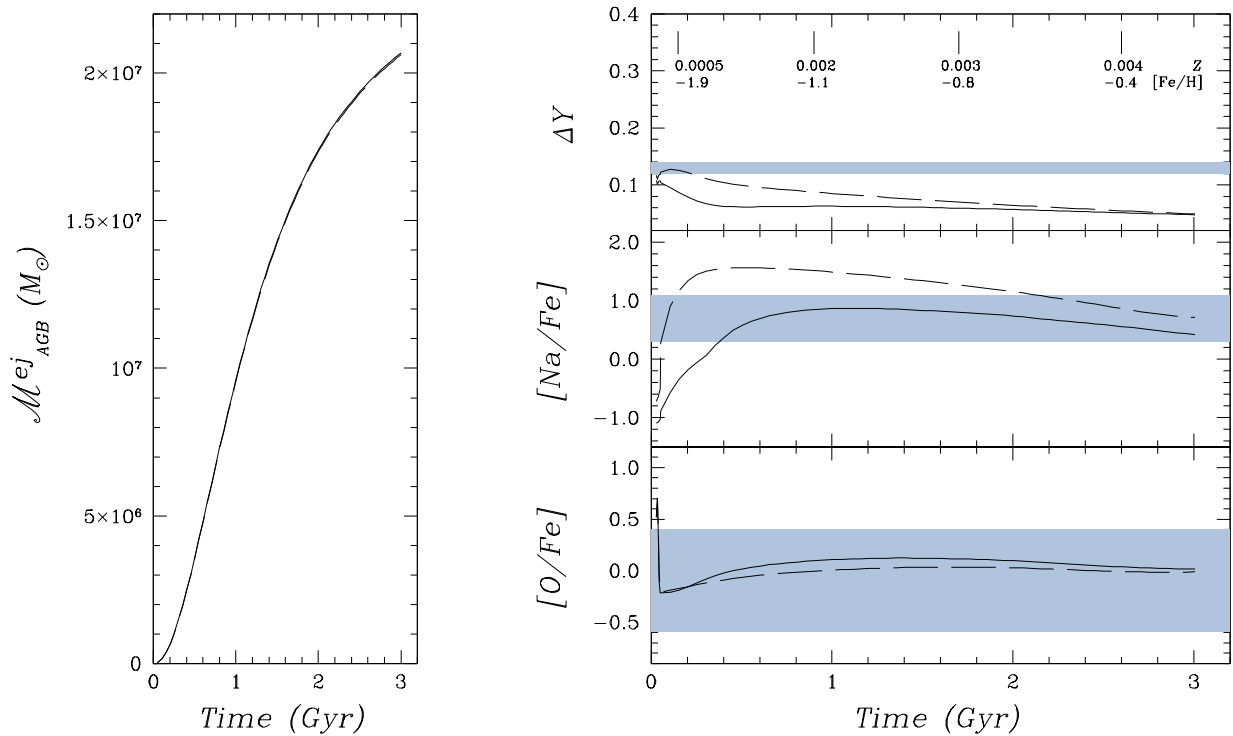}
     \caption{ Left panel: cumulative mass ejected by AGB stars in our model 
       for $\omega$\,Cen parent galaxy as a function of time. The chemical 
       composition of the ejecta as a function of time is shown in the 
       right-hand panels, for two sets of stellar yields (solid and dashed 
       lines; see text for details). The shaded areas in each of the right-hand 
       panels indicate observed values for chemically peculiar stars 
       (references in the text).}
     \label{fig:epAGB}
   \end{figure*}


   While it is reasonable to assume that Na is restored to the ISM through slow 
   stellar winds by both massive and AGB stars (and, hence, it is easily 
   retained within $\omega$\,Cen's progenitor potential well), oxygen, mostly a 
   product of He and C burning in massive stars, is released in the ISM by fast 
   radiatively-driven winds (Decressin et al. 2007a). Thus, it is easily lost 
   through the outflow. We assume an efficiency of O ejection through the 
   outflow of nearly 12 times the star formation efficiency. The same value 
   holds for all the ${\bmath{\alpha}}$-elements. It becomes slightly higher 
   for Fe (mostly produced by type Ia SNe) and the other Fe-peak elements: 13 
   times the star formation efficiency is the value which best fits both the 
   observed stellar metallicity distribution function and several abundance 
   ratios. This naturally leads to a Na-O anticorrelation in our model, as can 
   be seen from Fig.~\ref{fig:anticorr}, where we compare the trend of [Na/Fe] 
   versus [O/Fe] predicted by Model C (for four different assumptions on the 
   efficiency of Na entrainment in the outflow) with the available data. The 
   temporal evolution of the system can be read on the {\it x}-axis from right 
   to left. Although a large spread is present in the data, the average [O/Fe] 
   ratio in $\omega$\,Cen decreases with increasing metallicity, i.e. with time 
   in our models. We expect thus O-rich, Na-poor stars in $\omega$\,Cen to be 
   older than O-poor, Na-rich ones. This is at variance with models for 
   Galactic GCs by Marcolini et al. (2009), where the first stars to form are 
   Na-rich and O-depleted and `normal' (i.e. Na-depleted, O-rich) stars form 
   later. We are instead in agreement with Carretta et al.'s (2009) 
   interpretation of similar data for several Galactic globulars. Carretta et 
   al. (2009) suggest that O-rich, Na-depleted stars form out of gas of 
   `primordial' chemical composition, where `primordial' means the level of 
   chemical enrichment determined by the first episode of star formation. The 
   anticorrelation is then built up as long as stars of lower and lower initial 
   mass start to die and pollute the ISM with products of H burning at high 
   temperatures (see also Gratton et al. 2004).

   Our models, despite some simplifying assumptions, account satisfactorily 
   well for \emph{the trends of average abundances and abundance ratios} with 
   metallicity (see also Romano et al. 2007 and Romano \& Matteucci 2007) and 
   for observational constraints such as the Na-O anticorrelation in 
   $\omega$\,Cen. An important point to be stressed here is that also \emph{the 
     relative fractions of stars with normal and peculiar chemical 
     compositions} can be reproduced (see Sect.~\ref{sec:code}). The key 
   ingredient for the model predictions to be in agreement with the 
   observations is the development of a strong differential outflow in a much 
   more massive parent system. This outflow must vent out a major fraction of 
   the metals ejected by SN explosions. In turn, elements restored to the ISM 
   through gentle winds by both LIMSs and massive stars must be retained in the 
   shallow potential well of the cluster progenitor.

   In the following sections, we discuss the role of possible self-pollution 
   from either AGB or massive stars.


   \begin{figure*}
     \includegraphics[width=\textwidth]{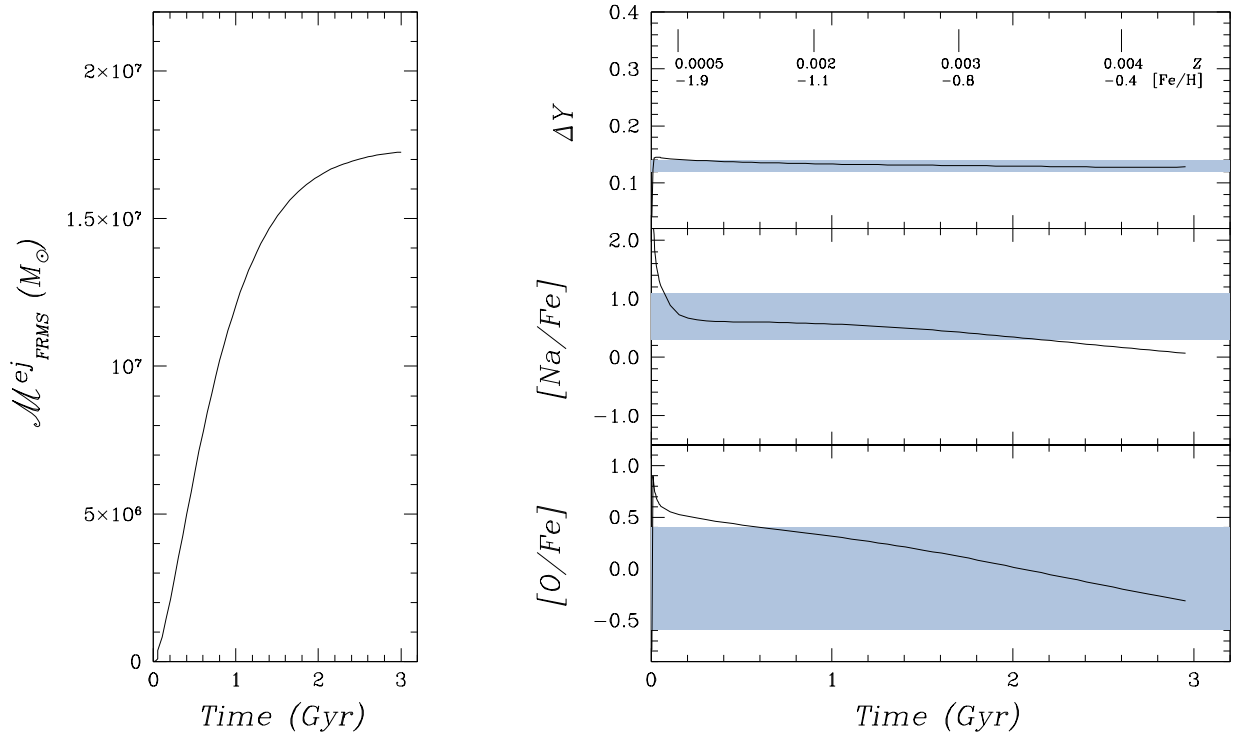}
     \caption{ Left panel: cumulative mass ejected in the slow winds by FRMSs 
       in the framework of our model for $\omega$\,Cen parent galaxy as a 
       function of time. The chemical composition of the ejecta as a function 
       of time is shown in the right-hand panels. The shaded areas in each of 
       the right-hand panels indicate observed values for chemically peculiar 
       stars (see text for references). Also shown in the upper part of the top 
       right panel is the run of the ISM metallicity with time for the 
       proto-$\omega$\,Cen host.}
     \label{fig:epMASS}
   \end{figure*}


   \subsubsection{AGB star pollution scenario}
   \label{sec:heAGB}

   Early claims that chemically peculiar GC stars could have formed from gas 
   processed in the envelopes of first-generation massive AGB stars (Cottrell 
   \& Da Costa 1981; Ventura et al. 2001) have recently been reconsidered and 
   given further support by detailed hydrodynamical and {\it N-}body 
   simulations (D'Ercole et al. 2008). These models assume the first stellar 
   generation already in place and start the calculations when all the SNeII 
   have already exploded. Since only stars in the 4-5 to 8~M$_{\odot}$ mass 
   range provide abundances in the ejecta compatible with the observations 
   (Ventura \& D'Antona 2008a,b), the second-generation stars must form in a 
   relatively short period of time -- 100~Myr in D'Ercole et al.'s model -- 
   before stars with masses $<$~4-5 M$_{\odot}$ can contribute to the chemical 
   enrichment. In the most extreme scenario, second-generation stars form out 
   of the AGB processed gas with an efficiency of 100 per cent and an IMF 
   completely skewed towards low-mass stars (only stars in the mass range 
   0.1--0.8 M$_{\odot}$ are allowed to form). This minimizes the mass of 
   first-generation stars which is needed in order to reproduce the fraction of 
   chemically peculiar stars currently observed in GCs. Under these conditions, 
   in fact, a cluster with a current mass $\mathscr{M}_{\mathrm{GC}}$ had a 10 
   times larger progenitor, while the mass of the progenitor must be much 
   larger if one or both of the aforementioned hypotheses is relaxed. This is 
   the case of our chemical evolution model for $\omega$\,Cen, where gas is 
   turned into stars with an efficiency of nearly 40 per cent and the stars 
   distributed according to a Salpeter-like IMF over the whole stellar mass 
   range (0.1--100 M$_\odot$; see the Appendix).

   In our standard model, $\omega$\,Cen's precursor is treated as a one-zone 
   system where the ejecta from massive stars, AGB stars (if any) and SNeIa (if 
   any) mix all together at each time. A contribution from infall of gas of 
   primordial chemical composition is considered as well. According to 
   Eq.(\ref{eq:infall}), the infall term exponentially decreases in time.

   To examine the `pure AGB' scenario, let us now assume that at a given time 
   -- or, better, time interval -- the ejecta of AGB stars stop mixing with the 
   surrounding medium and collect into the cluster core in a cooling flow (for 
   a detailed discussion on how cooling flows develop see D'Ercole et al. 
   2008). In Fig.~\ref{fig:epAGB}, left panel, we show the cumulative ejecta of 
   AGB stars belonging to the field stellar populations of $\omega$\,Cen's 
   parent galaxy as a function of time. The chemical composition of the ejecta 
   as a function of time is shown in the right-hand panels (solid and dashed 
   lines). In the top right panel we show the relative He enrichment, in the 
   middle right panel the [Na/Fe] ratio and in the bottom right panel the 
   [O/Fe] ratio. Also shown in the upper part of the top right panel is the run 
   of metallicity in the ISM of $\omega$\,Cen's progenitor with time. The 
   dashed lines refer to model predictions obtained by using the yields for AGB 
   stars by Karakas \& Lattanzio (2007), and the solid lines refer to model 
   predictions obtained by using the Karakas \& Lattanzio (2007) yields for AGB 
   stars below 3~M$_\odot$ and the Ventura \& D'Antona (2008a,b) yields for AGB 
   stars in the range 3--6.3~M$_\odot$. The ranges of $\Delta Y$, [Na/Fe] and 
   [O/Fe] values inferred from observations of chemically peculiar stars are 
   indicated by the shaded areas. Two important things are immediately apparent 
   from this figure. The first is that, in the framework of our model, AGB 
   stars provide nearly 2~$\times$ 10$^7$~M$_\odot$ of processed material. It 
   suffices that 10 per cent of this matter is converted into stars before any 
   mixing with the surrounding ISM to produce (assuming a standard Salpeter 
   IMF) about 10$^6$~M$_\odot$ of long-lived stars with a chemical composition 
   typical of pure AGB ejecta. The second important thing is that AGB stars 
   hardly produce the huge helium amount required to explain the blue MS of 
   $\omega$\,Cen (see also Karakas et al. 2006; Choi \& Yi 2008). The necessary 
   helium overabundance is attained only during the very early phases of the 
   proto-cluster evolution; then, as soon as stars less massive than 
   $\simeq$5~M$_\odot$ start to die, the overall AGB ejecta start to display 
   much more `normal' He abundances. Values of the oxygen-to-iron ratio as low 
   as [O/Fe]~= $-$0.5 are never reached for the same reason, while it seems 
   easier to get enhanced [Na/Fe] ratios\footnote{Notice that the Na yields 
     published by Karakas \& Lattanzio (2007) that we use here 
     (Fig.~\ref{fig:epAGB}, dashed lines) are being revised downwards 
     (A.~Karakas 2009, private communication). This should bring the model 
     predictions in better agreement with the observations.}. In the simplified 
   situation of two-population clusters, it may well be that only AGB stars 
   with initial masses in the range 4--5 to 8~M$_\odot$ have contributed to the 
   building up of second-generation stars. However, $\omega$\,Cen suffered a 
   much more complex evolutionary path (as witnessed by several indicators) and 
   we do not have any reason to exclude low-mass AGB stars from contributing to 
   the chemical enrichment. Rather, the exceptionally high abundances of 
   \emph{s-}process elements displayed by some cluster stars do suggest that 
   low-mass polluters played a fundamental role in the proto-cluster enrichment 
   (Johnson et al. 2009, and references therein).

   \subsubsection{Massive star pollution scenario}
   \label{sec:heMASS}

   In an attempt to overcome the shortcomings of the AGB star pollution 
   scenario, Decressin et al. (2007a,b) have proposed the winds of FRMSs as the 
   source of the chemical `anomalies' in GCs. Their study is limited to the 
   case of two-population clusters where the first generation of stars is 
   already in place.

   Here we analyse the implications of the winds of FRMSs scenario in the 
   framework of our complete modelling of the chemical enrichment of the whole 
   stellar population of $\omega$\,Cen. Of course, our analysis does not 
   correspond exactly to their case, because the original scenario strictly 
   applies to two-population clusters, whereas we are dealing with a much more 
   complex population in $\omega$\,Cen. In Fig.~\ref{fig:epMASS} we show the 
   cumulative ejecta in the winds of FRMSs during $\omega$\,Cen's parent galaxy 
   evolution (left panel), as well as their chemical composition as a function 
   of time: the He enrichment, [Na/Fe] ratio and [O/Fe] ratio are displayed in 
   the top right panel, middle right panel and bottom right panel, 
   respectively. The solid lines are the model predictions, while the shaded 
   areas represent the range of observed values for peculiar stars. It is seen 
   that any star formed from the pure ejecta of the slow winds from FRMSs will 
   have the extreme He abundance required by the location of the blue MS on the 
   CMD of $\omega$\,Cen, but the corresponding Na abundance will be too low. In 
   this respect we note that, while the yields of He and CNO elements from 
   FRMSs are provided for a wide grid of initial stellar metallicities (see 
   works by Meynet \& Maeder 2002; Hirschi et al. 2005; Hirschi 2007; Ekstr\"om 
   et al. 2008), Na yields have been computed only for stars with initial 
   metallicity [Fe/H]~= $-$1.5 (Decressin et al. 2007a). The adoption of 
   metallicity-dependent yields of Na from FRMSs could, in  principle, 
   counterbalance the effect on the [Na/Fe] ratio due to the increasing Fe 
   production from SNeIa with increasing time in $\omega$\,Cen host galaxy. Na 
   yields from massive stars computed by Kobayashi et al. (2006) without taking 
   stellar rotation into account increase with increasing the initial 
   metallicity of the stars (see Kobayashi et al. 2006, their table~3). Should 
   the inclusion of rotation preserve the trend found by Kobayashi et al. 
   (2006), Na production from FRMSs could eventually overwhelm that of Fe from 
   SNeIa. In this case, the O-poor stars born in the latest stages of the 
   proto-cluster formation (Fig.~\ref{fig:epMASS}, bottom right panel) would 
   also be Na-rich, thus producing the Na-O anticorrelation, as observed.

   Notice that the trend of decreasing [O/Fe] with time in the slow-wind ejecta 
   of FRMSs simply reflects the evolution of the [O/Fe] ratio in the ISM of the 
   host galaxy. Any O and Fe produced within the massive stars themselves, in 
   fact, do not enter the composition of the slow-wind ejecta. Both these 
   elements, in fact, are expelled later through fast polar winds and SN 
   explosions (see Decressin et al. 2007a).

   \section{Discussion and conclusions}
   \label{sec:end}

   It is common wisdom that $\omega$\,Cen is the naked nucleus of a dwarf 
   spheroidal galaxy that was ingested and partly disrupted by the Milky Way 
   some 10 Gyr ago (see e.g. Dinescu et al. 1999; Gnedin et al. 2002; Bekki \& 
   Norris 2006; Romano et al. 2007; Bellazzini et al. 2008). It is also 
   customarily acknowledged that metal enriched winds are the most 
   straightforward explanation of the observed properties of dwarf galaxies 
   (see the recent review by Tolstoy, Hill \& Tosi 2009, and references 
   therein).

   In the case of $\omega$\,Cen's progenitor we have found, indeed, that 
   significant outflow is needed in order to reduce the effective yield per 
   stellar generation and explain the observed properties (the stellar 
   metallicity distribution function, age-metallicity relation and the trends 
   of several abundance ratios with metallicity; Romano et al. 2007; see also 
   Ikuta \& Arimoto 2000). Following the results of hydrodynamical simulations 
   by Recchi et al. (2001), we have assumed that elements produced by SNe are 
   lost more efficiently than others. The lowest efficiency of ejection through 
   the outflow has been assigned to hydrogen and helium. This is a common 
   choice when dealing with the chemical evolution of local dwarf spheroidals 
   and it leads to a good reproduction of several observed properties (e.g. 
   Lanfranchi \& Matteucci 2003). However, no direct measurements of He are 
   available for local dwarf spheroidals: since they lack gas, they cannot be 
   observed for optical/radio recombination lines. Thus, the efficiency of He 
   entrainment in the galactic wind is, as a matter of fact, totally 
   unconstrained. With this in mind, the challenging bet is: could 
   $\omega$\,Cen's parent galaxy \emph{retain most of its helium,} while still 
   \emph{ejecting most of the heaviest species?}

   Up to now, three major scenarios have been proposed for the origin of the 
   helium-enriched populations in $\omega$\,Cen, as well as in other massive 
   GCs:
   \begin{enumerate}
     \item[A] accretion of helium-rich material by pre-existing stars;
       \item[B] star formation out of the ejecta from either massive AGB stars 
         (B1) or fast rotating massive stars (B2);
         \item[C] pollution by Pop\,{\sevensize III} stars.
   \end{enumerate}
   Recently, Renzini (2008) has reviewed them critically and concluded that 
   only the AGB option (B1) appears to be acceptable; the others result in 
   either a spread of helium abundances (A and B2) or a certain degree of 
   enrichment in heavy elements as well (C). However, for the AGB option to 
   work, AGB stars more massive than 3~M$_{\odot}$ must experience just a few 
   TDU episodes and the cluster hosting the He-rich population must be the 
   remnant of a more massive systems. Detailed stellar modelling (Ventura \& 
   D'Antona 2008a,b) shows that only AGB stars above 4--5~M$_{\odot}$ can 
   actually reach the required levels of He enhancement in their atmospheres.

   In this paper we have thoroughly analysed the issue of the formation of 
   He-rich stars in $\omega$\,Cen by means of a complete model for its chemical 
   evolution. In $\omega$\,Cen we are facing a \emph{primary population} (about 
   50 per cent of the stars, with $\langle$[Fe/H]$\rangle \simeq -$1.7 and 
   `normal' He) and a \emph{secondary population} (at intermediate, 
   $\langle$[Fe/H]$\rangle \simeq -$1.4, and extreme, $\langle$[Fe/H]$\rangle 
   \simeq -$0.6, metallicities). At least part of the secondary-population 
   stars may be He enhanced.

   Although the interaction with the Milky Way is likely to have played some 
   role in shaping the chemical properties of (at least part of) the stellar 
   population in $\omega$\,Cen, we find a good agreement between predictions 
   from standard chemical evolution models (that do not take all the relevant 
   dynamical processes into account) and observations of average abundances and 
   abundance ratios by assuming that a galactic-scale outflow vented out mainly 
   matter enriched in SN products, while mostly retaining the ejecta of slow 
   stellar winds, irrespective of the initial mass of the stellar progenitor. 
   Besides this ejection of enriched gaseous matter, the cluster progenitor 
   must have lost a major fraction of stars somehow later on during its 
   evolution. In summary, we show that, in order to explain the existence of 
   extreme He-rich stellar populations, \emph{what really matters is the kind 
     of evolution the host system went through, rather than the kind of stars 
     responsible for a `super He production'.}

   In $\omega$\,Cen, because of the relatively long-lasting star formation 
   (some 10$^9$ years), stars of initial mass as low as 2~M$_{\odot}$ had the 
   time to contribute significantly to the chemical enrichment of the ISM. This 
   is clearly witnessed by the high \emph{s-}process abundances displayed by 
   some cluster stars (e.g. Johnson et al. 2009). If cooling flows developed to 
   collect the AGB ejecta in the innermost galactic regions where chemically 
   peculiar stars were born, it can be hardly told which selective mechanism 
   brought only the ejecta from 4--5 to 8~M$_\odot$ AGB stars to the cluster 
   core, while leaving behind those from lower-mass stars. Including a 
   contribution to the chemical enrichment from super-AGB stars could help the 
   AGB star pollution scenario to produce results in agreement with the 
   observations (e.g. Pumo et al. 2008), but up to now no nucleosynthetic 
   yields from this class of objects have been provided in the literature for 
   use in chemical evolution studies.

   The competing scenario of pollution from slow winds of FRMSs, while being 
   able to reproduce, in principle, the peculiar abundances and abundance 
   ratios required by the observations of chemically peculiar stars, is 
   hampered by a number of assumptions about, for instance, the efficiency of 
   meridional circulation to mix the products of core H-burning in the stellar 
   envelope, or the rate of mass loss through the mechanical wind, or the need 
   for the development of a large contingent of fast rotators in GCs. Since 
   the matter expelled through the equatorial discs of FRMSs is available with 
   the `right' He abundance since the very beginning of the proto-$\omega$\,Cen 
   formation (see Sect.~\ref{sec:heMASS} and Fig.~\ref{fig:epMASS}), it is not 
   clear why an extreme He-rich population should form only later on, at 
   metallicities around [Fe/H]~$\sim -$1.2 (and possibly higher).

   As a last comment, it is worth stressing that abundances of He as high as 
   that required to explain the blue MS of $\omega$\,Cen have never been 
   observed elsewhere. Although we may think about mechanisms able to originate 
   extreme He-rich stellar populations, we must also be aware that the 
   existence of such extremely high He abundances still await a confirmation 
   from spectroscopic observations.

   \section*{Acknowledgments}

   DR thanks Eugenio Carretta and Angela Bragaglia for enlightening 
   conversations on the chemical composition of globular cluster stars and 
   Antonio Sollima for constructive comments. We are grateful to Paolo Ventura 
   and Amanda Karakas for providing their yields in a friendly format, and to 
   the anonymous referee for suggestions that improved significantly the 
   clarity of the paper. MC, FM, DR and MT acknowledge partial financial 
   support from italian MIUR through grant PRIN~2007, prot.~2007JJC53X\_001.

\appendix

   \section{Model basic assumptions and equations}
   \label{sec:basic}

   The chemical evolution model is one zone, with instantaneous and complete 
   mixing of gas inside it. The instantaneous recycling approximation is 
   relaxed, i.e. the stellar lifetimes are taken into account in detail. We 
   follow the evolution of several stable chemical species and their isotopes 
   (H, D, He, Li, C, N, O, F, Na, Mg, Al, Si, S, K, Ca, Sc, Ti, V, Cr, Mn, Co, 
   Fe, Ni, Cu, Zn) by means of integro-differential equations taking the form
   \begin{eqnarray}
     \lefteqn{ \frac{{\mathrm{d}}{\mathscr{G}}_i(t)}{{\mathrm{d}}t} = 
       -X_i(t)\psi(t) }
     \nonumber \\
       & & {}+\int_{m_{l}}^{m_{B_{m}}} \!\! \psi(t - 
       \tau_{m}){\mathscr{Q}}_{mi}(t - \tau_{m})\varphi(m) {\mathrm{d}}m 
       \nonumber \\
       & & {}+A \int_{m_{B_{m}}}^{m_{B_{M}}} \!\! \varphi(m_B)
       \nonumber \\
       & & {}\cdot \int_{\mu_{\mathrm{min}}}^{0.5} \!\! f(\mu) \psi(t - 
       \tau_{m_2}) {\mathscr{Q}}_{m_1i}(t - \tau_{m_2}) 
       {\mathrm{d}}\mu \ {\mathrm{d}}m_B 
       \nonumber \\
       & & {}+(1 - A) \int_{m_{B_{m}}}^{m_{B_{M}}} \!\! \psi(t - \tau_{m})
       {\mathscr{Q}}_{mi}(t - \tau_{m})\varphi(m) {\mathrm{d}}m
       \nonumber \\
       & & {}+\int_{m_{B_{M}}}^{m_{u}} \!\! \psi(t - 
       \tau_{m}){\mathscr{Q}}_{mi}(t - \tau_{m})\varphi(m) {\mathrm{d}}m 
       \nonumber \\
       & & {}+\frac{{\mathrm{d}}{\mathscr{G}}_{i}^{\mathrm{in}}(t)}
       {{\mathrm{d}}t} 
       -\frac{{\mathrm{d}}{\mathscr{G}}_{i}^{\mathrm{out}}(t)}{{\mathrm{d}}t}.
       \label{eq:bas}
   \end{eqnarray}
   Eq.(\ref{eq:bas}) shows that the abundance of a given element $i$ changes 
   with time because of the processes of star formation, mass return from dying 
   stars and infall (outflow) of gas towards (from) the system.

   In particular, ${\mathscr{G}}_i(t)$ is the gaseous mass in form of element 
   $i$ at a given time $t$ normalized to a barionic mass 
   ${\mathscr{M}}_{\mathrm{bar}}$~= 10$^9$ M$_\odot$ (see Sect.~\ref{sec:code}), 
   $X_i(t)$ is the abundance by mass of element $i$ at the time $t$, $\psi(t)$ 
   is the star formation rate (SFR) at the time $t$, $\tau_m$ is the lifetime 
   of stars with initial mass $m$ and $\varphi(m)$ is the IMF.

   The SFR is expressed as:
   \begin{equation}
     \psi(t) = \nu {\mathscr{G}}^k(t),
   \end{equation}
   where $\nu$~= 0.35 Gyr$^{-1}$ is the star formation efficiency, 
   ${\mathscr{G}}(t)$~= 
   ${\mathscr{M}}_{\mathrm{gas}}(t)/{\mathscr{M}}_{\mathrm{bar}}$ is the 
   normalized gaseous mass at the time $t$ and $k$~= 1.

   At variance with Galactic field populations, current observational evidence 
   seems to favour a Salpeter-like IMF in both dwarf spheroidals and Galactic 
   globulars (Wyse 2005, and references therein). Hence, we assume a Salpeter 
   (1955) IMF, extrapolated and normalized to unity over the 
   0.1--100~M$_{\odot}$ stellar mass range. However, the current mass function 
   of $\omega$\,Centauri is better reproduced by a broken power law, with 
   indices $\alpha$~= $-$2.3 for $m >$ 0.5~M$_{\odot}$ and $\alpha$~= $-$0.8 
   for $m <$ 0.5~M$_{\odot}$ (Sollima, Ferraro \& Bellazzini 2007). We have 
   checked that the results presented in this paper are not affected by our 
   choice of the IMF, because the low-mass stars act just as an extra sink of 
   matter in chemical evolution models.

   The integrals on the right hand side of Eq.(\ref{eq:bas}) represent the rate 
   at which element $i$ is restored to the interstellar medium (ISM) by dying 
   stars. We adopt the production matrix formalism (Talbot \& Arnett 1973) and 
   compute the quantities
   \begin{equation}
   {\mathscr{Q}}_{mi}(t - \tau_{m}) = {\mathscr{Q}}_{m}X_i(t - \tau_{m}),
   \end{equation}
   i.e. the fractional mass of the star of initial mass $m$ that is restored to 
   the ISM as element $i$ when the star dies, according to different 
   nucleosynthesis prescriptions. The dependence on time is driven by the 
   dependence of the yields on metallicity. The integration limits $m_l$ and 
   $m_u$ refer to the lowest (0.8 M$_\odot$) and highest (100 M$_\odot$) mass 
   contributing to galactic enrichment, while $m_{B_{m}}$~= 3 M$_\odot$ and 
   $m_{B_{M}}$~= 16 M$_\odot$ are the lower and upper limits for the total mass 
   of binary systems leading to Type Ia supernova (SNIa) explosions (Matteucci 
   \& Greggio 1986). The first, third and fourth integrals refer to the 
   contributions from single stars; the second one takes into account the 
   chemical enrichment by SNeIa: $m_B$ is the total mass of the system giving 
   rise to a SNIa explosion, $m_1$ and $m_2$ are the masses of the primary and 
   secondary stars, respectively, $f(\mu)$ is the distribution function for the 
   mass fraction of the secondary (after Matteucci \& Greggio 1986, to which we 
   refer the reader for more details). $A$ is a free parameter, constant in 
   time and space. It gives the fraction of mass per stellar generation that 
   ends up in SNIa precursors. It is fixed by the SNIa rates observed in 
   galaxies of different morphological type.

   The last two terms on the right hand side of Eq.(\ref{eq:bas}) represent the 
   positive contribution from infall of pristine matter and the negative one of 
   any outflow of gas from the system. They read:
   \begin{equation}
     \frac{{\mathrm{d}}{\mathscr{G}}_{i}^{\mathrm{in}}(t)}{{\mathrm{d}}t} =
     \frac{X_i^{\mathrm{in}} \exp{(-t/\tau)}}{\tau[1 - 
       \exp{(-t_{\mathrm{now}}/\tau)}]}
     \label{eq:infall}
   \end{equation}
   and
   \begin{equation}
     \frac{{\mathrm{d}}{\mathscr{G}}_{i}^{\mathrm{out}}(t)}{{\mathrm{d}}t} = 
     w_iX_i(t){\mathscr{G}}(t),
   \end{equation}
   respectively, where $\tau$~= 0.5 Gyr is the time-scale for infall, 
   $X_i^{\mathrm{in}}$ are the abundances of the infalling gas (set to their 
   primordial values; see Romano et al. 2003), $t_{\mathrm{now}}$~= 13.7 Gyr 
   (Tegmark et al. 2006) is the age of the universe now and $w_i$ (in units of 
   Gyr$^{-1}$) is a free parameter that describes the efficiency of the 
   galactic wind (it takes different values for different elements; in 
   particular, it takes into account the results of dynamical studies that SN 
   ejecta leave the galaxy more easily than the unperturbed ISM -- see, e.g., 
   Recchi, Matteucci \& D'Ercole 2001, and references therein). The time of the 
   onset of the galactic wind is self-consistently computed (see Romano et al. 
   2007, and references therein).

\bsp

\label{lastpage}

\end{document}